\documentstyle[twocolumn,psfig,prl,aps]{revtex} 
\begin{document}
\draft
\title{Integral Equation Results for the $^4$He(e,\,e$^\prime$p)$^3$H 
       Reaction at High Missing Momenta}
\author{S. A. Sofianos, G. Ellerkmann}
\address{Physics Department, University of South Africa,  
         P.O.Box 392,Pretoria 0003, South Africa}
\author{W. Sandhas} 
\address{Physikalisches Institut, Universit\"{a}t 
         Bonn, D-53115 Bonn, Germany}  
\date{\today}                     
\maketitle
\begin{abstract} 
The two-fragment electrodisintegration of $^4$He into proton and triton
is calculated in Plane Wave Impulse Approximation (PWIA). 
The three- and four-nucleon wave functions involved
are obtained by solving the Alt-Grassberger-Sandhas (AGS) 
integral equations, with the Malfliet-Tjon potential
as the underlying NN-interaction. Our results are in remarkable 
agreement with the experimental data and, in contrast to alternative
approaches, do not exhibit any dip in the five-fold differential 
cross section at a missing momentum of~$\sim$~450\,MeV/$c$.\\
\end{abstract} 
%
%

The \, two-fragment \, electrodisintegration \, process
$^4$He(e,\,e$^\prime$p)$^3$H has been the  subject of several 
experimental investigations for various kinematics (see, for example,
\cite{Brand88,Brand91,Goff,Leeuth,Leeuwe98}). On the theoretical side
quite some effort has been devoted to calculating this process.
However, the exact treatment of four-nucleon electrodisintegration observables
is computationally very demanding and, thus, has usually been simplified
by approximations and model assumptions
 ~\cite{Leeuwe98,Schia90,Laget94,Howell97}. 

In Plane-Wave Impulse Approximation (PWIA) all these calculations 
exhibit a characteristic dip, actually zero, in the five-fold 
differential cross-section
around a missing momentum of  $\sim$~450\,MeV/$c$, which does not show up
in the  experimental data. Laget \cite{Laget94} performed calculations 
including final state interaction (FSI) effects and meson exchange 
currents (MEC) by means of a Feynman diagrammatic approach. Although
this resulted in a partial filling of the dip, these investigations 
also underestimate the data considerably in this region. 
Similar results were obtained when FSI was taken into account via an 
effective nucleon-trinucleon interaction \cite{Schia90,Howell97,Sand98}.
In a completely different approach Nagorny {\it et al.}\cite{Nagorny}
included the electromagnetic field within the strongly interacting 
system in a relativistic gauge invariant way, the FSI being 
incorporated via the pole contribution of the
p{}$^3$H$\rightarrow$ p{}$^3$H  scattering matrix \cite{Zay}.
The agreement with the data is again fairly satisfactory,  but the zero 
is exhibited as well \cite{Leeuwe98}. 

In detail, at missing momenta less than 300\,MeV/$c$ all 
calculations show a
good agreement with the data \cite{Brand91,Leeuwe98,Howell97}.
Surprisingly,  
the PWIA performs reasonably well in this region where the FSI could be 
expected to be more important than in the higher missing momenta
region. In contrast, in the region $300$\,MeV/$c$\,$<Q<600$\,MeV/$c$  
the results strongly depend on the way the FSI effects are included.  
For example, as pointed out in  \cite{Leeuwe98}  the Laget results 
underestimate the cross section by a factor of 4 and those
of Schiavilla  by a factor of 2.
At missing momenta above 600\,MeV/$c$, where the MEC contribution
is becoming important, the agreement with the data is again fair. 
We, therefore, conclude that the zero in the PWIA cross section is not 
necessarily a manifestation of strong FSI or MEC effects. 

Instead, one should look for other explanations, such as the dependence
of the results on the model used, the NN forces employed, the
determination  of the bound state wave functions etc. In the 
present work the wave functions involved were calculated within the exact 
three- and four-nucleon AGS formalism \cite{ags3,ags4}.  We mention that
the same wave functions have successfully been employed already
in calculations of the two-fragment photodisintegration of 
the $\alpha$--particle \cite{ell1,ell2}.

%
We consider the two-fragment reaction  
in which the scattered electron and the ejected nucleon are
measured  in coincidence. The  corresponding electron-proton 
coincidence cross section is given by 
\begin{equation} 
        \frac{{\rm d}^5\sigma}{{\rm d}E_f d\Omega_p {\rm d}\Omega_e} =  
                \frac{\sigma_{\rm M}}{(\hbar 
                c)^3 (2\pi)^3} 
                \frac{\rho_f}{4E_iE_f\cos^2
\displaystyle{\frac{\theta}{2}}}  
                |{\cal M}({\bf q})|^2 
\label{tcross} 
\end{equation} 
where $\sigma_{\rm M}$ is the Mott differential cross section,  
\begin{equation} 
        \sigma_{\rm M} = \frac{e^4\cos^2  
        \displaystyle{\frac{\theta}{2}}}{4 E_i^2 \sin^4 
        \displaystyle{\frac{\theta}{2}}}\,. 
\end{equation} 
$E_i(E_f)$ is the energy of the incoming (outgoing) electron and
$\rho_f$ is the relativistic density of states. 
The transition matrix, properly antisymmetrized with respect to 
the four nucleons \cite{boe80}, is given by  
\begin{equation} 
        {\cal M}({\bf q}) = 2 \;
        ^{(-)}\langle{\bf q};\Psi_{III}|H|\Psi_{IV}\rangle\,,  
\label{tmatrix} 
\end{equation} 
where $H$ is the Hamiltonian describing the 
interaction between the electron and the nucleons. The ejected 
proton moves away with momentum ${\bf q}$ with respect  to the 
residual three-nucleon bound state $|\Psi_{III}\rangle$.
The kinematics of this process is shown in Fig. 1. 

The  Hamiltonian for the interaction between an electron 
and four nucleons is that of McVoy and van Hove \cite{McVoy}, which 
has been previously employed in the electrodisintegration of the 
trinucleon system  by Lehman and collaborators \cite{Lehman_I,Lehman}
and by Epp and Griffy \cite{Epp}. 
This Hamiltonian, correct to the order of  $\hbar^2 Q^2/M^2c^2$,
is   
\begin{eqnarray} 
        H & = & -\frac{4\pi e^2}{q_\mu^2}\langle v_f|\sum_{j=1}^4 
         \left\{F_{1N}(q_\mu^2)~e^{-i {\bf Q} \cdot 
         {\bf x}_j} \phantom{\frac{q^2}{8M}}  
          \right.\nonumber \\  
        & - & \frac{F_{1N}(q_\mu^2)}{2M} [({\bf p}_j \cdot 
        \mbox{\boldmath  $\alpha$})~ e^{-i {\bf Q}  
        \cdot {\bf x}_j} +   
           e^{-i{\bf Q} \cdot {\bf x}_j}~ 
         ({\bf p}_j \cdot \mbox{\boldmath $\alpha$})] 
         \nonumber\\  
           & - & i \left[ \frac{F_{1N}(q_\mu^2) + \kappa 
         F_{2N}(q_\mu^2)}{2M}\right] 
         \mbox{\boldmath $\sigma$}_j\cdot ({\bf x}_j \times 
        \mbox{\boldmath  $\alpha$})~  
         e^{-i {\bf Q} \cdot {\bf x}_j} \nonumber\\ 
        & + & \left. \frac{q_\mu^2}{8M^2}~[F_{1N}(q_\mu^2) + 2\kappa 
         F_{2N}(q_\mu^2)]~  
          e^{-i {\bf Q} \cdot {\bf x}_j} \right \} 
        |u_i\rangle\,.
        \label{hamiltonian} 
\end{eqnarray} 
Here ${\bf x}_j$ and ${\bf p}_j$  are the position and momentum 
operators of the $j$-th  nucleon,  ${\mbox{\boldmath $\sigma$}}_j$ 
is the nucleon spin operator, ${\mbox{\boldmath $\alpha$}}$ is the 
Dirac matrix acting on the free electron spinors 
$|v_i\rangle $ and $|v_f\rangle$,  while $q_\mu^2$  is the exchanged 
four-momentum  squared. $F_{1N}$ and $F_{2N}$ are the 
form factors of the nucleon, $\kappa$ is the 
anomalous moment of the nucleon in nuclear magnetons, and $M$ is the 
nucleon mass.

For   proton knock-out, the transition matrix  Eq.~(\ref{tmatrix}) 
reads
\begin{equation} 
        {\cal M} = - \langle v_f| v_i\rangle {\cal M}_Q  
        + \langle v_f|{\mbox{\boldmath $\alpha$}}|v_i\rangle   
        \cdot \left ( {\bf M}_{{\rm el}} 
        +{\bf M}_{{\rm mag}}\right)\,, 
\end{equation} 
where 
\begin{eqnarray} 
        {\cal M}_Q & = &2 \,{}^{(-)}\langle {\bf q}; 
        \Psi_{III}| {\cal H}_Q    |\Psi_{IV} \rangle\,,  \label{Mq}\\ 
        {\bf M}_{\rm el} & = &2 {}^{(-)}\langle {\bf q}; \Psi_{III}|
        {\bf H}_{{\rm el}}   |\Psi_{IV} \rangle \,, \label{Mel}  \\ 
        {\bf M}_{\rm mag} & = &2 {}^{(-)}\langle {\bf q}; \Psi_{III}|  
        {\bf H}_{{\rm mag}}|\Psi_{IV} \rangle \,.\label{Magn}  
\end{eqnarray} 
The Hamiltonians ${\cal H}_Q$, ${\bf H}_{{\rm el}}$, and ${\bf H}_{{\rm
mag}}$ 
are given by 
\begin{eqnarray} 
\label{h1} 
        {\cal H}_Q & = & F_{\rm ch}^p(1+q_\mu^2/8M^2) \, \sum_{j=1}^4 
        e^{-i {\bf Q} \cdot {\bf x}_j}\,\lambda_j \,,\\ 
\label{hel} 
        {\bf H}_{{\rm el}} & = & (F_{\rm    ch}^p/2M) \, \sum_{j=1}^4  
        ({\bf p}_j e^{-i {\bf Q} \cdot 
        {\bf x}_j}+e^{-i {\bf Q} \cdot {\bf x}_j} {\bf p}_j)           
        \,\lambda_j\,, 
\\ 
\label{hmag} 
        {\bf H}_{\rm mag} & = & (i/2M)F_{\rm mag}^p 
        \sum_{j=1}^4 \,  e^{-i {\bf Q} \cdot {\bf x}_j} 
        \mbox{\boldmath $\sigma$}_j\times{\bf Q} \lambda_j\,, 
\label{jel} 
\end{eqnarray} 
Here the superscript $p$ refers to the proton and   
$\lambda_j=(1+\tau_z^j)/2$ is the isospin operator for nucleon $j$
while $F_{\rm ch}^p$ and $F_{\rm mag}^p$ are the  charge and magnetic 
form factors of the proton defined by 
\begin{eqnarray} 
        F_{\rm ch}^p & = & F_{1p} + (q_\mu^2/4M^2)\kappa_p F_{2p} \\ 
        F_{\rm mag}^p & = & F_{1p} + \kappa_p F_{2p}\,. 
\end{eqnarray} 
The analytical fit to the proton form factors $F_{1p}$ and $F_{2p}$ 
given by Janssens {\it et al.} \cite{Janssens} is used in the
calculations.                                              

Squaring the matrix element,  summing and averaging over the electron 
spin, and inserting the resulting expression in Eq.~(\ref{tcross}),
we obtain 
\begin{eqnarray} 
        \frac{d^5\sigma}{dE_f\,d\Omega_p\,d\Omega_e} & = 
        &\frac{\sigma_{\rm M}}{(\hbar c)^3 (2\pi)^3}~ 
        \frac{|{\bf p}_p|E_p}{1 - \displaystyle 
          \frac{E_p}{E_{^3 H}}   \frac{{\bf p}_p  
        \cdot {\bf p}_{^3 H}}{|{\bf p}_p|^2}} 
\nonumber \\  \left\{|{\cal M}_Q|^2 \right.  
        &-& \frac{1}{2} \sec^2 \frac{\theta}{2}
           ({\cal M}_Q^*    {\bf J}+{\bf J}^*{\cal M}_Q)  
        \cdot(\hat{k}_i + \hat{k}_f)  
\nonumber \\ 
        &+& \frac{1}{2} \sec^2 \frac{\theta}{2} ({\bf J}\cdot \hat{k}_i
        {\bf J}^* \cdot \hat{k}_f + {\bf J}\cdot \hat{k}_f 
        {\bf J}^* \cdot \hat{k}_i) 
\nonumber \\  
        &+& \left. |{\bf J}|^2 \tan^2 \frac{\theta}{2}\right\}\,, 
        \label{xsec} 
\end{eqnarray} 
where 
        ${\bf J}={\bf M}_{{\rm el}} +{\bf M}_{{\rm mag}} $.
The determination of the coincidence cross section is thus reduced to
the 
determination of the nuclear matrix elements ${\cal M}_Q$ and ${\bf J}$.
                      
In this work we use the dominant electric operators (\ref{h1}) and 
(\ref{hel}). In  PWIA the nuclear matrix elements (\ref{Mq}) and
(\ref{Mel}) 
read
\begin{equation} 
        {\cal B}_Q({\bf q}) = 2 \, \langle {\bf q}|\langle \Psi_{III}| 
          \sum_{j=1}^4 \, \exp(-i {\bf Q}  
         \cdot{\bf x}_j)\,\lambda_j 
         |\Psi_{IV}\rangle  
         \label{bornq} 
\end{equation} 
and 
\begin{eqnarray} 
        {\bf B}_{{\rm el}} ({\bf q}) &=& 2 \, \langle{\bf q}|\langle
\Psi_{III}| 
          \sum_{j=1}^4 \,\left( 
        {\bf p}_j  
        \exp(-i {\bf Q}\cdot{\bf x}_j) \right.
\nonumber\\ 
        & + & \left. \exp(-i {\bf Q}\cdot{\bf x}_j) 
        {\bf p}_j\right) 
        \,\lambda_j  |\Psi_{IV}\rangle  \ ,
        \label{bornj} 
\end{eqnarray} 
where 
        ${\bf q} = ({\bf p}_1 + {\bf p}_2 + {\bf p}_3 - 3 {\bf p}_4)/4$.

The operators appearing in Eqs. (\ref{bornq}) 
and (\ref{bornj})  are the same as those of Eqs. (\ref{h1}) and
(\ref{hel}), 
except that the nucleonic form factors  $F_{\rm ch}^p$ and 
$F_{\rm mag}^p$ are not noted here. 

To proceed we express the operators  
${\bf x}_j$ and ${\bf p}_j$ in Jacobi coordinates and neglect, as in 
the photodisintegration case \cite{ell1,ell2}, those acting 
within $|\Psi_{III}\rangle$. The remaining terms, containing ${\bf q}$
and its canonically conjugate counterpart, are treated without further
approximation. A straightforward calculation then reduces (\ref{bornq})
to 
\begin{eqnarray} 
     {\cal B}^{\prime}_Q({\bf q})& = & 2 \,\langle {\bf q} + \frac{1}{4}
{\bf Q}|\langle \Psi_{III}|  
     (\lambda_1 + \lambda_2 + \lambda_3)
              |\Psi_{IV}\rangle \nonumber \\ 
     & + & 2 \, \langle {\bf q} -\frac{3}{4} {\bf Q}|\langle \Psi_{III}|
     \lambda_4
     |\Psi_{IV}\rangle , 
     \label{bornprexq} 
\end{eqnarray}                 
whereas (\ref{bornj}) is replaced by
\begin{eqnarray} 
     {\bf B}^{\prime}_{{\rm el}} ({\bf q})& = & \frac{4}{3} \, {\bf q}
\,
         \langle {\bf q}+\frac{1}{4} {\bf Q}|\langle \Psi_{III}|  
     (\lambda_1 + \lambda_2 + \lambda_3)
              |\Psi_{IV}\rangle \nonumber \\ 
     & - & 4 \, {\bf q} \,
         \langle {\bf q}-\frac{3}{4} {\bf Q}|\langle \Psi_{III}|  
     \lambda_4   
         |\Psi_{IV}\rangle  + \, {\bf Q} \, {\cal B}^{\prime}_Q({\bf
q}). 
     \label{bornprexj} 
\end{eqnarray} 
The construction of the above matrix elements requires the knowledge 
of the bound states $|\Psi_{III}\rangle$ and $|\Psi_{IV}\rangle$. 
For their calculation the exact three- and four-nucleon AGS integral equations
are  employed \cite{ags3,ags4}. The latter consist the coupled set of  
(before antisymmetrization) 18x18 
four-body AGS equations. They contain in their kernel all subsystem
information via the two-body T-matrices, the three- and (2+2)-body AGS
transition operators. By this approach, the full coupling and the corresponding 
interference of (2+2)- and (3+1)-channels in the four-body system 
is taken into account explicitly, and thus exactly and completely. 
In order to reduce the original three-
and four-body relations to (one-dimensional) integral equations, 
the W-matrix method \cite{wm86} and the energy-dependent 
pole approximation (EDPA) \cite{edpe78} are used. In the purely 
nuclear case these approximations have led to very accurate
results (see e.g. \cite{wmc1,wmc2,edpe}). Furthermore, they have been 
successfully used in  calculations of the photodisintegration of 
$^3$H, $^3$He \cite{Sand98,Schadow} and $^4$He
\cite{ell1,ell2,Sand98}. 
The graphical representation of matrix elements 
like Eqs. (\ref{bornq}) and (\ref{bornj}),
adapted to the four-body AGS formalism, can be found in \cite{boe80}.
As in \cite{ell1,ell2}, the Malfliet-Tjon potential I and III
\cite{mt6970}
is chosen, as it is both sufficiently realistic and simple enough to 
be employed in four-nucleon calculations. This property is of particular
importance in our calculations where the computation of the 
matrix elements, despite the approximations used,
is still tedious and of considerable numerical complexity.
The corresponding binding energies are 8.595 MeV for $^3$H 
and 30.1 MeV for $^4$He \cite{ell1}.
 
The results obtained within the AGS formalism for the
$^4$He(e,e$^\prime$p)$^3$H
five-fold differential cross section as a function of the missing
momentum ${\bf Q}$  are shown in Fig. 2.  The kinematics and the
experimental data are those of \cite{Leeuth,Leeuwe98} for the
$\omega=215$\,MeV case. For comparison we also included in the figure 
the PWIA results of \cite{Howell97}, obtained for wave functions
constructed via the integrodifferential equation approach (IDEA) 
of Ref. \cite{IDEA}. The PWIA results of Laget (see Refs. 
\cite{Leeuth,Leeuwe98,Laget94}) are
also shown, being obtained for the Urbana potential and for wave
functions constructed with the variational Monte Carlo (MC) method. 
The agreement of our AGS calculations with the experimental data, 
especially in the region 
where the PWIA results of the other two methods show their 
characteristic dip, is remarkable. This holds true also in comparison
with the other results reported in \cite{Leeuwe98}.
Fig. 3 shows the five-fold differential cross section  
for the  {\rm $^4$He(e,e$^\prime$p)$^3$H} reaction for the Saclay
kinematics \cite{Goff}. For comparison the Laget results 
\cite{Laget94} are also plotted.
The agreement of our results with  experiment is again remarkable. 

The overall small discrepancies 
may be reduced  by using a 
better NN force, further improvements in the PWIA matrix elements,
and  inclusion of the  FSI in a rigorous way. A particular 
advantage of the AGS-type approach in this respect
lies in the fact that the incorporation of the FSI is exactly the 
same for the four-nucleon scattering, the photodisintegration of
$^4$He, and the electrodisintegration of $^4$He (see e.g. 
\cite{Sand98} and Refs. therein). 
Most important: in this approach the 
underlying integral equations 
explicitly incorporate the (2+2)-channels, 
not fully included in other approaches,
and their coupling to the (3+1)-channels. 
That underlying interference of 
competing channels is the most relevant feature of four-body 
theory as compared to three-body theory. 
In other words,
the complexity of four-body rearrangement processes is
fully taken into account.

In conclusion, our results show that already in PWIA quite a good 
description of the experimental data can be achieved. The main reason
for this agreement appears to be the use of wave functions obtained from 
the AGS integral equations with their complete coupling scheme. 
Another reason is the way of calculating the 
nuclear matrix elements. Namely, those parts of 
the electromagnetic operators (\ref{h1}) and (\ref{hel}), which act 
between the relative motion of the two outgoing nuclear fragments, are 
taken into account exactly. The sensitivity to the input NN-potential 
and to the above-mentioned 2+2 rearrangement terms are 
under investigation.

\acknowledgements
Financial support from the  University of South
Africa  and the Foundation for Research Development of South
Africa is appreciated.


%
\begin{figure}[htb]
\centerline{\psfig{file=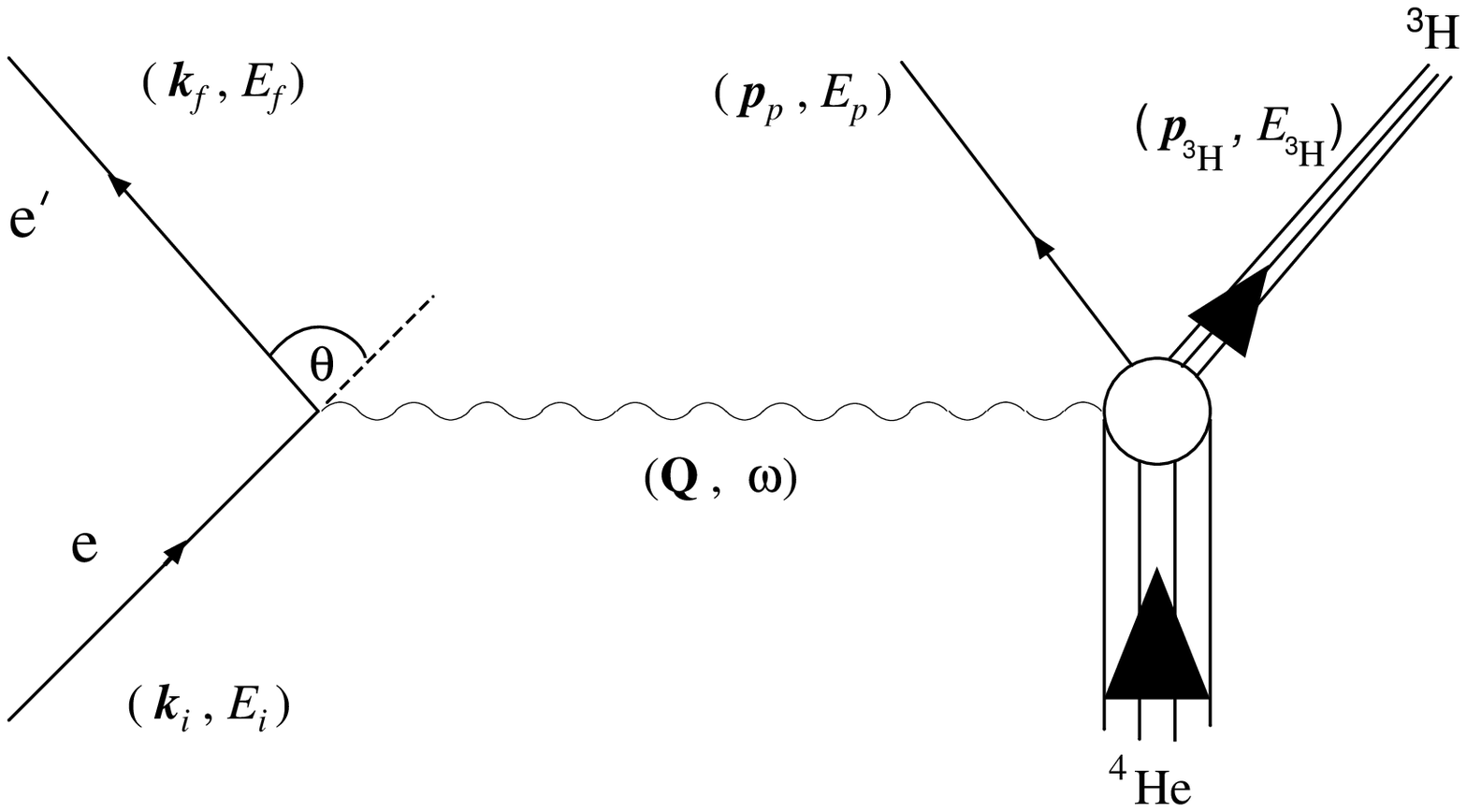,height=6cm,width=8cm}}
 \caption{\label{fig1} 
     Kinematics for the process ${\rm {}^4He(e,e^\prime p){}^3H}$.}
\end{figure}
\begin{figure}[htb]
\centerline{\psfig{file=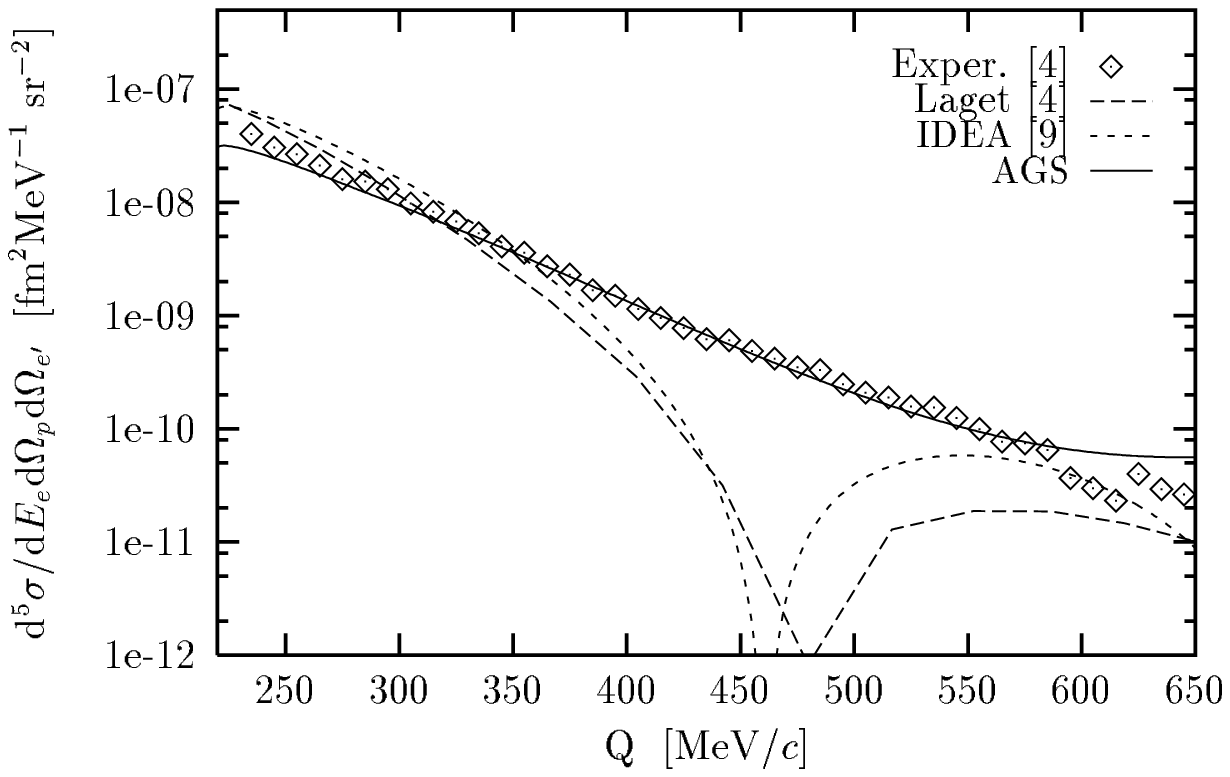,height=6cm,width=8cm,angle=-360}}
 \caption{\label{fig2} Five-fold {\rm $^4$He(e,e$^\prime$ p)$^3$H} 
differential cross section as a function of the missing 
momentum {\rm Q} for the  NIKHEF kinematics [4,5].}

\end{figure}
\begin{figure}[htb]
\centerline{\psfig{file=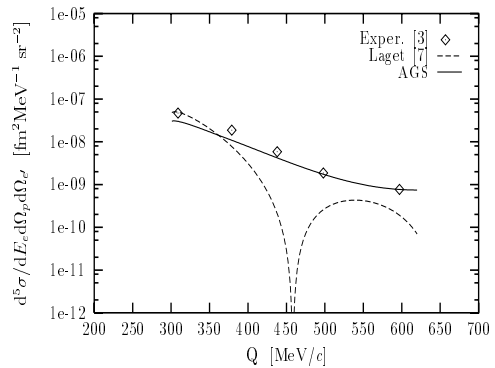,height=6cm,width=8cm,angle=360}}
\caption{\label{fig3} Five-fold  {\rm $^4$He(e,e$^\prime$p)$^3$H}  
differential cross section  as a function of the missing 
momentum  {\rm Q}  for the the Saclay kinematics [3].} 
\end{figure}
\end{document}